\documentclass[]{rnaastex}

\usepackage{graphicx}
\usepackage[suffix=]{epstopdf}
\usepackage{natbib}
\usepackage{amsmath}
\usepackage{xspace}

\usepackage{hyperref}

\begin{document}



\newcommand{\kepler}{\emph{Kepler}}

\newcommand{\python}{{\tt PYTHON}}
\newcommand{\phasma}{{\tt phasma}}
\newcommand{\forecaster}{{\tt forecaster}}
\newcommand{\emcee}{{\tt emcee}}
\newcommand{\scipy}{{\tt scipy}}
\newcommand{\astropy}{{\tt astropy}}
\newcommand{\sinc}{\mathrm{sinc}}

\title{Two Thousand \kepler\ Phase Curves from \phasma}

\correspondingauthor{David Kipping}
\email{dkipping@astro.columbia.edu}

\author{David Kipping, Emily Sandford \& Tiffany Jansen}
\affiliation{Department of Astronomy, Columbia University, 550 W 120th Street, New York NY}


\section{} 

An exoplanet's optical phase curve constrains the thermal emission and albedo
of the planet's surface and/or atmosphere (e.g. see \citealt{kipping:2011,
demory:2013}), as well as potentially constraining the mass via gravitational
influences on the host star (e.g. see \citealt{borucki:2009,barclay:2012}).
Recently, \citet{jansen:2018} demonstrated that exoplanets with precisely
constrained orbital periods, as is typical of transiting planets, enable one to
exploit a Fourier-argument to non-parametrically separate optical phase curves
from long-term photometric time series in the presence of realistic noise
structures. This algorithm, dubbed \phasma, was applied to 477 exoplanets in
\citet{jansen:2018}, with the final phase curves made publicly available at
\href{https://github.com/CoolWorlds/phasma}{github.com/CoolWorlds/phasma}.

\phasma\ works best when the baseline of each uninterrupted photometric time
series is much greater than the orbital period. Since each \kepler\ quarter
is $\sim90$\,days in duration, we limited our scope to $P<10$\,d period
planets. In addition, we required that planets had a NASA Exoplanet Archive
(NEA) disposition of ``CONFIRMED'' and a parent star with $\log g>4$. Here, we
provide the community with a larger set of detrended \phasma\ phase curves
by relaxing these latter two constraints. The surface gravity constraint is
removed altogether. The dispostion required is relaxed to anything for
which either the \kepler\ or NEA disposition does not equal ``FALSE POSITIVE'',
excepting cases where either disposition is ``CONFIRMED''. Our processing
exactly follows the methods described in \citet{jansen:2018} otherwise,
using \kepler\ DR25 PDC data throughout.

The new sample contains 1,998 KOIs, including the original 477, and we make
the detrended lightcurves and summary figures (example shown in Figure 1)
publicly available at Columbia's Academic Commons
\dataset[(DOI: 10.7916/D8SB5NR7)]{https://doi.org/10.7916/D8SB5NR7}. Each
summary figure includes a solid-black line representing a double harmonic
($P$ and $P/2$) regressed to the points occuring outside of transit and
occultation, which should not be treated as a physical fit but rather a
means for guiding the eye when intepretting the figures.

In addition, we have recoded the source code into \python\ to enable the
community to implement the package for themselves more easily on K2 and TESS
light curves in the future; available at
\href{https://github.com/CoolWorlds/phasma}{github.com/CoolWorlds/phasma}.

\phasma\ provides a powerful non-parametric detrending approach with
zero-tuning parameters able to compete with the precision obtained by much more
human-intensive approaches (e.g. see comparison to \citealt{armstrong:2016} in
\citealt{jansen:2018}). These results and software should enable the community
to more easily work with phase curve science products from ongoing surveys.

\begin{figure*}
\begin{center}
\includegraphics[width=16.0cm,angle=0,clip=true]{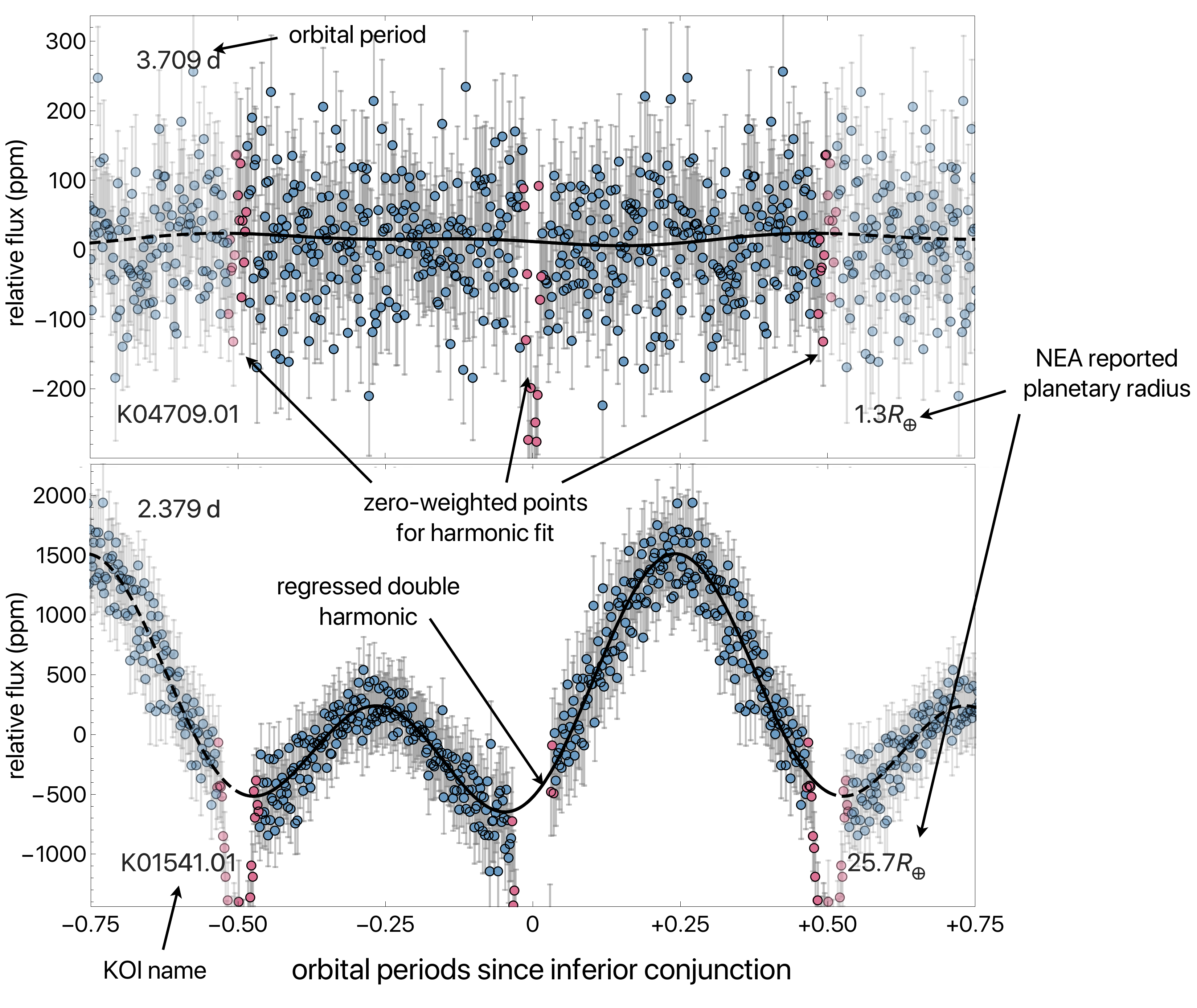}
\caption{\emph{
Two examples of the summary figures automatically generated by us for each KOI.
Lower case is a typical strong detection whereas the upper case is effectively
flat.
}}
\label{fig:1}
\end{center}
\end{figure*}

\acknowledgments

DMK is supported by the Alfred P. Sloan Foundation.


\end{document}